\documentclass[aps, prd, showpacs, superscriptaddress, nofootinbib, twocolumn]{revtex4}
\usepackage{amssymb,amsmath,color,microtype}
\usepackage{graphicx}

\def\beq{\begin{equation}}
\def\eeq{\end{equation}}
\def\bea{\begin{eqnarray}}
\def\eea{\end{eqnarray}}

\begin{document}

\title{The Effect of Primordial Non-Gaussianities on the Seeds of Super-Massive Black Holes}

\author{Zeinab Sherkatghanad}
\email{z.sherkat@mail.mcgill.ca}
\affiliation{Department of Physics, McGill University, Montr\'eal, QC, H3A 2T8, Canada}

\author{Robert H. Brandenberger}
\email{rhb@hep.physics.mcgill.ca}
\affiliation{Department of Physics, McGill University, Montr\'eal, QC, H3A 2T8, Canada}

\begin{abstract}

The origin of the seeds which develop into the observed super-massive black holes 
at high redshifts may be hard to interpret in the context of the standard $\Lambda CDM$ of 
early universe cosmology based on Gaussian primordial perturbations. Here we consider 
the modification of the halo mass function obtained by introducing skewness and kurtosis 
of the primordial fluctuations. We show that such primordial non-Gaussianities constrained
by the current  observational bounds on the nonlinearity parameters of $f_{NL}$ and $g_{NL}$ 
are not effective at greatly increasing the number density of seeds which could 
develop into super-massive black holes at high redshifts. This is to be contrasted
with the role which cosmic string loops could play in seeding super-massive black holes.

\end{abstract}


\maketitle
 \section{Introduction}\label{sec:intro}
 
Super-massive black holes (SMBHs) with masses exceeding $10^6 M_\odot$ ($M_\odot$ 
denotes the solar mass) have recently attracted a lot of attention 
\cite{Alexander, Wu, Fan, Jiang, Marconi, Willott, Banados, Volonteri-Stark, Menou, Pollack, micic, Begelman, Bromm}. 
It is now believed 
that each galaxy contains at least one super-massive black hole which forms from the 
accretion of gas about massive seed objects. The origin of the seeds which cause the 
formation of SMBHs is still somewhat of a mystery \cite{Volonteri,Johnson}. According 
to the standard paradigm of early universe in which the primordial cosmological fluctuations 
are approximately Gaussian and have an almost scale invariant spectrum, nonlinearities 
form only at fairly late times and there may not be enough time to produce the nonlinear 
massive seeds which seed SMBHs of mass greater or equal to $10^9 M_\odot$ at redshifts 
of $6$ or higher (of which more than $40$ candidates are now known \cite{Willott, Banados}).
There are three types of candidate seeds for SMBHs, namely Population III stars with masses
in the range between $10^2 M_\odot$ and $10^3 M_\odot$, dense matter clouds with 
masses between $10^3 M_\odot$ and $10^6 M_\odot$, and compact objects of mass 
between $10^2 M_\odot$ and $10^4 M_\odot$ formed by the collision of old stellar clusters.

Recently \cite{Robert} it has been shown that cosmic string loops which result from a 
scaling solution of strings formed during a phase transition in the very early universe lead to 
an additional source of compact seeds. The number density of string-induced seeds 
dominates at high redshifts and can help trigger the formation of the observed super-massive 
black holes. Cosmic string loops form a special type of non-Gaussian density fluctuation field. 
Non-Gaussianities in the primordial density perturbation field are an inevitable consequence of the
nonlinearities of the Einstein field equations. However, in minimal models of matter and in
the context of an inflationary origin of the density fluctuations
such non-Gaussianities have a very small amplitude. Larger amplitudes can be obtained
in non-minimal models of inflation \cite{non-minimal} and in some alternatives to
inflation such as the ``Ekpyrotic'' scenario \cite{Ekp} and in the 
``matter bounce'' \cite{Wands}. In such models, the non-Gaussianities are usually
parametrized in terms of the skewness and kurtosis which characterize the deviations
from Gaussianity in the three and four point functions. Skewness and kurtosis are
a good characteristic of non-Gaussianities emerging from a distortion of originally
Gaussian fluctuations. In this paper we ask whether such non-Gaussianities can
play a similar role as cosmic string loops in the triggering of the formation of
nonlinear seeds for SMBHs at high redshifts.
 
Primordial perturbations which originate as quantum fluctuations 
of a scalar field and which seed cosmic microwave background (CMB) fluctuations and structure 
formation of the Universe lead to non-Gaussianities which are well characterized by
skewness and kurtosis (in contrast to string loop-induced fluctuations which 
are intrinsically non-Gaussian and hence not well characterized by the three
and four point functions only. Nevertheless, the conventional non-Gaussianities 
resulting from a deformation of an initially Gaussian process 
can yield information about the early stages of the evolution of our Universe. In
particular, they can allow discrimination between various models of inflation and their 
alternatives. 

In general, the three and four point functions are characterized by
an amplitude and a shape. Here, we will focus on local type non-Gaussian perturbations 
which can be written in the following form \cite{Salopek,Wang}
\beq \label{non}
\zeta \, = \, \zeta_G + \frac{3}{5} f_{NL} (\zeta_G ^2 -\langle\zeta_G ^2\rangle)
+ \frac{9}{5} g_{NL} \zeta_G ^3.
\eeq
where $\zeta$ is the primordial curvature fluctuation variable, $\zeta_G$ is the Gaussian part 
and $f_{NL}$ and $g_{NL}$ are nonlinearity parameters. These nonlinearity parameters 
$f_{NL}$ and $g_{NL}$ lead to non zero values of the skewness and kurtosis of the primordial 
non-Gaussianities in the Probability Density Function (PDF), respectively 
\cite{Silk, Desjacques, Smith, Maggiore}. Current observations 
constrain the magnitude of the parameters $f_{NL}$ and $g_{NL}$ \cite{PlanckCollaboration}
to be $f_{NL}^{equil} = - 16 \pm 79$ and $g_{NL}^{local} = (-9 \pm 7.7) \times 10^4$ (both
at 68\% confidence level). Here, the superscripts stand for the shape, ``equilateral'' in the
first case and ``local'' in the second.

In this paper we take into acount the modification of the halo mass function in the presence of 
skewness and kurtosis of local type (following the method of \cite{Amico, Silk})
and determine the effects of these primordial non-Gaussianities 
on the mass of the dark matter halos at different redshifts (up to $z=20$) compared with
the situation in a purely Gaussian model. We demonstrate that the constraints which come 
from the current observational data on the magnitude of the nonlinearity parameters are already
strong enough to prevent the non-Gaussianities from eliminating the difficulties that the 
standard $\Lambda  CDM$ of early universe cosmology may face in terms of explaining
the origin of the observed super-massive black holes at high redshifts. 

The paper is organized as follows. First we start with a short review of the halo mass function in 
the $\Lambda CDM$ model. In Section III we study the modifications of the halo mass function 
resulting from introducing non-vanishing skewness and kurtosis of the primordial density perturbations.
We evaluate the result for skewness and kurtosis of equilateral shape. We show that 
this type of primordial non-Gaussianities obeying the current observational bounds on 
the nonlinearity parameters $f_{NL}$ and $g_{NL}$ is not effective at greatly enhancing
the number density of nonlinear seeds for SMBHs at high redshifts. We conclude with
a discussion section.

\section{Halo Mass function in the $\Lambda CDM$ Model}

The linear relative matter density fluctuation can be written in terms of the primordial curvature 
fluctuations $\zeta(k)$ on uniform energy density hypersurfaces by \cite{structurebook,Silk}
\beq
\delta(\vec{k},z) \, = \, \frac{2 k^2}{5 \Omega_{m0} H_0^2} T(k) D(z) \zeta(\vec{k})~,
\eeq
where ${\vec{k}}$ denotes comoving wavenumber, $z$ is the cosmological
redshift, $\Omega_{m0}$ is the present density parameter, $H_0$ is the Hubble constant, 
$D(z)$ is a linear growth function and $T(k)$ is a transfer function.
Hence, the linear matter power spectrum $P_{\delta}(k)$ is given by the curvature power 
spectrum $P_{\zeta}$ via
\bea
\langle\delta(\vec{k}_1,z) \delta(\vec{k}_2,z)\rangle \, &=& \, 
(2 \pi ^3) P_{\delta}(k_1,z) \delta ^3 (\vec{k}_1 +\vec{k}_2) \nonumber \\
P_{\delta}(k_1,z) \, &=& \, {\cal{A}}(k_1,z) ^2 P_{\zeta}(k_1)~,
\eea
where 
\beq
{\cal{A}}(k,z) \, = \, \frac{2 k^2}{5 \Omega_{m0} H_0^2} T(k) D(z) \, .
\eeq
 
The smothed density fluctuation on a given length, $R$, is defined by
\beq
\delta_R \, = \, \int \frac{d^3 \vec{k}}{(2 \pi)^3} W_R (k) \delta(\vec{k},z)~,
\eeq
with $W_R (k)$ being the Fourier transform of a window function
\beq
W_R (k) \, = \, \frac{3 \ (\sin kR-kR \cos kR)}{(kR)^3}~.
\eeq
The variance  in mass on a momentum scale $k$ with a top-hat filter with a radius 
that encloses the mass is (for an infinite total spatial volume) given by
\bea
\sigma_R^2 \, &=& \, \int \frac{d k}{k} W_R^2 (k) \  {\cal{A}}(k,z)^2 {\cal{P}}_{\zeta}(k) \nonumber \\
&=& \, \int \frac{d k}{k} W_R^2 (k) \  {\cal{P}}_{\delta}(k)~,
\eea
where the dimensionless power spectra ${\cal P}$ are related to the dimensional ones
$P$ via 
\bea
{\cal{P}}_{\zeta}(k) \, &=& \, \frac{k^3}{2 \pi^2} P_{\zeta}(k) \, = \, A_s (\frac{k}{k_s})^{n_s-1} \nonumber \\
{\cal{P}}_{\delta}(k) \, &=& \, \frac{k^3}{2 \pi^2} P_{\delta}(k) \, .
\eea 

The halo mass function based on the Press-Schechter theory \cite{PS} is given by \cite{Sheth}
\bea\label{halo}
\frac{dn}{dM}(M,z) dM \, &=& \, -dM \frac{2 \rho}{M} \frac{d}{dM} \int _{\delta_c/\sigma_R} ^{\infty} d\nu F(\nu)  
\\
&=& \, -dM \sqrt{\frac{2}{\pi }} \frac{\rho}{M}  \exp [-\frac{\nu_c^2}{2}] \ \nu_c \ \frac{d \log \sigma_R}{d M} ~,
\nonumber
\eea
and
\beq
 d\nu F(\nu) \, = \, \frac{d\nu}{\sqrt{2 \pi}} \exp(-\frac{\nu^2}{2})~,
\eeq
where $\rho$ is the mean mass density of the universe (background density) and 
$\nu_c=\frac{\delta_c}{\sigma_R}$. The number $\delta_c$ is the threshold for collapse, 
and we will use the value  $\delta_c=1.86$ which corresponds to neglecting the effects of dark energy.

According to linear perturbation theory, the density contrast $\sigma_R$ grows linearly in
the scale factor before the contribution of dark energy to the equation of state of matter becomes
important. The above formulas then show that for a Gaussian spectrum of fluctuations, the
number density of nonlinear seeds falls off exponentially at high redshifts, thus making it
difficult to account for the origin of the nonlinear seeds which are needed to explain the
origin of SMBH. We will now investigate whether the addition of skewness and kurtosis 
can improve the prospects for high redshift SMBH formation.

\section{Modification of the halo mass function by Primordial Non-Gaussianities}

In this section we focus on the local type non-Gaussianities and study the modifications of 
the halo mass function when introducing primordial skewness and kurtosis
(see also \cite{Matarrese, LoVerde} for other studies of the effects of
non-Gaussianities on structure formation). We take
the primordial curvature fluctuations to be given by (\ref{non}) in terms of a
Gaussian distribution. Based on this relation the two, three and four point functions become
\bea\label{POTRBI}
\langle\zeta(\vec{k}_1) \zeta(\vec{k}_2)\rangle \, &=& \, 
(2 \pi)^3 P_{\zeta}(k_1) \delta^{(3)} (\vec{k}_1+\vec{k}_2) \\ \nonumber
\langle\zeta(\vec{k}_1) \zeta(\vec{k}_2) \zeta(\vec{k}_3)\rangle &=& 
(2 \pi)^3 \frac{6}{5} f_{NL} (P_{\zeta}(k_1) P_{\zeta}(k_2)\\\nonumber
&& +2\ perms.) \ \delta^{(3)} (\vec{k}_1+\vec{k}_2+\vec{k}_3) \\\nonumber
\langle\zeta(\vec{k}_1) \zeta(\vec{k}_2) \zeta(\vec{k}_3) \zeta(\vec{k}_4)\rangle &=& 
(2 \pi)^3 \frac{54}{25} g_{NL} (P_{\zeta}(k_1) \\\nonumber
&& P_{\zeta}(k_2) P_{\zeta}(k_3)+3 \ perms.) \\ \nonumber 
&& \delta^{(3)} (\vec{k}_1+\vec{k}_2+\vec{k}_3+\vec{k}_4)~.
\eea

In order to consider the effects of primordial non-Gaussianities on the smoothed density 
fluctuations, we can define the n-th central moment of the Probability Density Function (PDF)
$F(\delta_R) d\delta_R$ in the following standard way
\beq\label{DP}
\langle \delta _R ^n \rangle \, = \, \int_{-\infty} ^{\infty} \delta _R ^n F(\delta_R) d\delta_R~.
\eeq
In addition, we can define the reduced $p$-th cumulant $S_p$ in this form
\beq\label{cumulant}
S_p (R) \, = \, \frac{\langle \delta _R ^p \rangle_c}{\langle\delta _R ^2\rangle_c ^{p-1}}~, 
\eeq
where
\beq
\langle \delta _R  \rangle_c \, = \, 0  \  ,\  \langle \delta _R ^2 \rangle_c \, = \, \sigma_R ^2  \  ,    \  etc \, .
\eeq

If we take into acount a non-Gaussian PDF of matter density perturbations in Eq. (\ref{DP}), we 
can build up the PDF from the cumulants and the Gaussian distribution by using the 
Edgeworth expansion. This technique gives the non-Gaussian PDF of the density field in 
terms of derivatives of the Gaussian PDF and reduced cumulants \cite{Matarrese, LoVerde}
\beq \label{Edge1}
F(\nu) d\nu \, = \, F_G(\nu) \ d\nu+\sum_{m=3} \frac{c_m}{m!} F_G ^{(m)} (\nu) \ d\nu~,
\eeq
with
\bea \label{Edge2}
F_G (\nu) \, &=& \, \frac{1}{\sqrt{2 \pi}} exp(-\nu^2 / 2)\\ \nonumber
F_G ^{(m)} (\nu) \, &=& \,\frac{d^m}{d\nu ^m} F_G(\nu) \, = \, (-1)^m H_m(\nu)F_G (\nu)~,
\eea
where $F_G  (\nu)$ is the Gaussian PDF and $H_m(\nu)$ are the Hermite polynomials
\bea \label{Hermite}
H_1(\nu)&=&\nu, \  H_2(\nu)=\nu^2-1, \ H_3(\nu)=\nu ^3-3 \nu \nonumber \\
H_4(\nu)&=&\nu ^4-6 \nu^2+3 \ ,  \  ...... 
\eea
The coefficients $c_m$ are given by
\beq \label{coeff}
c_m \, = \, (-1)^m \int_{-\infty} ^{\infty} H_m(\nu) F(\nu) d\nu~,
\eeq
with
\beq \label{coeff2}
c_3 \, = \, -S_3(R) \sigma_R, \ , \ c_4 \, = \, S_4(R) \sigma_R^2 \ , \ ....
\eeq

This technique allows us to calculate the non-Gaussian PDF of the density field in 
terms of the non zero values of the skewness and kurtosis of the primordial 
fluctuations. Inserting (\ref{Edge2}) into (\ref{Edge1}) and making use of
(\ref{coeff2}) we obtain
\bea\label{PDF}
F(\nu) d\nu &=& \frac{d \nu}{\sqrt{2 \pi}} exp(-\nu^2 / 2) [1+\frac{S_3(R) \sigma_R}{6} H_3 (\nu)\\
\nonumber
&& +  \frac{1}{2} (\frac{S_3(R) \sigma_R}{6})^2 H_6 (\nu)+\frac{1}{6} (\frac{S_3(R) \sigma_R}{6})^3 H_9 (\nu)\\
\nonumber
&& + \frac{S_4(R) \sigma_R^2}{24} H_4 (\nu)+\frac{1}{2}(\frac{S_4(R) \sigma_R^2}{24})^2 H_8 (\nu)\\
\nonumber
&& + \frac{1}{6} (\frac{S_4(R) \sigma_R^2}{24})^3 H_{12} (\nu)+\ ... \ ]~,
\eea
where we are neglecting terms of higher order in $S_N(R)$
(from (\ref{POTRBI}), (\ref{DP}) and (\ref{cumulant})), and where
\bea
S_3(R) \, &=& \, \frac{6}{5}\frac{f_{NL}}{\sigma_R^4} \int \frac{d k_1}{k_1} W_R (k_1) \  
{\cal{A}}(k_1) {\cal{P}}_{\zeta}(k_1)\\
\nonumber
&& \times\int \frac{d k_2}{k_2} W_R (k_2) \   {\cal{A}}(k_2) {\cal{P}}_{\zeta}(k_2) \\
\nonumber
&& \times \int_{-1} ^{1} \frac{d \mu_{12}}{2} W_R (k_{12}) \  {\cal{A}}(k_{12}) \ 
\Big(1+\frac{P_{\zeta}(k_{12})}{P_{\zeta}(k_1)}+\frac{P_{\zeta}(k_{12})}{P_{\zeta}(k_2)}\Big)~,
\eea
and
\bea
S_4(R) \, &=& \, \frac{54}{25}\frac{g_{NL}}{\sigma_R^6} \int \frac{d k_1}{k_1} W_R (k_1) \  {\cal{A}}(k_1) {\cal{P}}_{\zeta}(k_1)\\
\nonumber
&& \times\int \frac{d k_2}{k_2} W_R (k_2) \   {\cal{A}}(k_2)  {\cal{P}}_{\zeta}(k_2)\\
\nonumber
&& \times \int \frac{d k_{3}}{2 \pi k_{3}} W_R (k_3) \   {\cal{A}}(k_{3})  {\cal{P}}_{\zeta}(k_{3})\\
\nonumber
&& \times\int_{-1} ^1 \frac{d \mu_{12}}{2}  \int_{-1} ^1 \frac{d \mu_{123}}{2} \int_{0} ^{2 \pi} d\phi_{13} \ W_R (k_{123}) \  {\cal{A}}(k_{123}) \\
\nonumber
&& \times \Big(1+\frac{P_{\zeta}(k_{123})}{P_{\zeta}(k_1)}+\frac{P_{\zeta}(k_{123})}{P_{\zeta}(k_2)}+\frac{P_{\zeta}(k_{123})}{P_{\zeta}(k_{3})}\Big)~,
\eea
and where 
\bea
\mu_{ij} \, &=& \, cos \ \theta _{ij}\, , \nonumber \\
k_{12} \, &=& \, \sqrt{k_1^2+k_2^2+2 k_1 k_2 \mu_{12}} \, {\rm and} \\ \nonumber
k_{123} \, &=& \, \Big[k_1^2+k_2^2+k_3^2+2 k_1 k_2 \mu_{12}+ 2 k_1 k_3 \mu_{13}\\\nonumber
&& +2 k_2 k_3\Big(\sqrt{(1-\mu_{12}^2)(1-\mu_{13}^2) \cos \phi_{13}+\mu_{12} \mu_{13}}\Big)\Big]^{1/2}~.
\eea

Thus, the modified halo mass function based on the Press-Schechter formula with a
non-Gaussian PDF of the smoothed density field is given by (from Eqs. (\ref{halo}) and (\ref{PDF}))
\bea
\label{halototal}
&\frac{dn}{dM}&(M,z) dM \, = \, -dM \frac{2 \rho}{M} \frac{d}{dM} \int _{\delta_c/\sigma_R} ^{\infty} d\nu F(\nu)\\
\nonumber
&=& -dM \sqrt{\frac{2}{\pi }} \frac{\rho}{M}  \exp [-\frac{\nu_c^2}{2}] \ \Bigg(\nu_c \frac{d \log \sigma_R}{d M} \ \Big[1\\
\nonumber
&& + \frac{S_3 \sigma_R}{6} H_3(\nu_c)+\frac{S_4 \sigma_R ^2}{24} H_4(\nu_c)\\\nonumber
&& + \frac{1}{2} H_6(\nu_c) \left(\frac{S_3 \sigma_R}{6}\right)^2+\frac{1}{2} H_8(\nu_c) \left(\frac{S_4 \sigma_R^2}{24}\right)^2\\\nonumber
&& + \frac{1}{6} H_9(\nu_c) \left(\frac{S_3 \sigma_R}{6}\right)^3+\frac{1}{6} H_{12} (\nu_c) \left(\frac{S_4 \sigma_R^2}{24}\right)^3\Big] \\\nonumber
&& + H_2(\nu_c) \frac{d}{d M}\left(\frac{S_3 \sigma_R}{6}\right)+H_3(\nu_c) \frac{d }{d M}\left(\frac{S_4 \sigma_R^2}{24}\right)\\\nonumber
&& + \frac{1}{2} H_5(\nu_c) \frac{d }{d M}\left(\frac{S_3 \sigma_R}{6}\right)^2+\frac{1}{2} H_7(\nu_c) \frac{d }{d M}\left(\frac{S_4 \sigma_R^2}{24}\right)^2\\\nonumber
&& + \frac{1}{6} H_8(\nu_c) \frac{d}{d M}\left(\frac{S_3 \sigma_R}{6}\right)^3 \\\nonumber
&& +\frac{1}{6} H_{11}(\nu_c) \frac{d }{d M}\left(\frac{S_4 \sigma_R^2}{24}\right)^3 \Bigg) + .....
\eea
here $\nu_c=\frac{\delta_c}{\sigma_R}$ and $\delta_c$ is the threshold for collapse. 

In what follows we consider two important forms of non-Gaussianity for cosmological observations,
namely the local form and the equilateral form.  The {\it local form} of the bispectrum requires 
that one of the three momentum modes exits the Hubble radius (e.g. in the context of
an inflationary universe) much earlier than the other two, 
i.e $k_1\ll k_2 \simeq k_3$. Taking this form of the non-Gaussianity for skewness and kurtosis 
we get
\bea
S_3(R) \, &=& \, \frac{12}{5}\frac{f_{NL}}{\sigma_R^2} \int \frac{d k_1}{k_1} W_R (k_1) \  
{\cal{A}}(k_1) {\cal{P}}_{\zeta}(k_1)~,
\eea
and
\bea
S_4(R) \, &=& \, \frac{162}{25}\frac{g_{NL}}{\sigma_R^4} \int \frac{d k_1}{k_1} W_R (k_1) \  
{\cal{A}}(k_1) {\cal{P}}_{\zeta}(k_1)  \nonumber \\
&& \times \int \frac{d k_2}{k_2} W_R (k_2) \  {\cal{A}}(k_2) {\cal{P}}_{\zeta}(k_2)~.
\eea

For the {\it equilateral} form of non-Gaussianity, i.e $k_1=k_2=k_3=k_4=k$, we have
\beq
\label{S34}
S_3(R) \, = \, \frac{18}{5}\frac{f_{NL}}{\sigma_R^2} \int \frac{d k}{k} W_R (k) \  
{\cal{A}}(k) {\cal{P}}_{\zeta}(k)
\eeq
and
\beq
S_4(R) \, = \, \frac{216}{25}\frac{g_{NL}}{\sigma_R^4} \Big(\int \frac{d k}{k} W_R (k) \  
{\cal{A}}(k) {\cal{P}}_{\zeta}(k)\Big)^2~.
\eeq

We can now estimate the magnitude of the skewness and kurtosis in the 
equilateral limit. Let us consider a sharp k-space filter which is particularly 
useful for theoretical  arguments and is the equivalent to the top-hat filter in 
Fourier space \cite{structurebook},
\bea
W_R (k) \, = \, \left\{
  \begin{array}{lr}
    1 & \  kR \le 1\\
    0 & \  kR>1
  \end{array}
\right.
\eea
In real space this takes the form
\bea
W_R (x) \, = \, \frac{3}{V}\ |\frac{x}{R}|^{-3} (\sin \frac{|x|}{R}-\frac{|x|}{R} 
\cos \frac{|x|}{R})~,
\eea
where the volume is taken to be $V=6 \pi^2 R^3$. Therefore, we obtain 
\beq
S_3(R) \sigma_R \, \simeq \, \frac{18}{5}f_{NL} \sqrt{{\cal{P}}_{\zeta}(1/R)}\simeq\frac{0.8 }{5}\times 10^{-3} f_{NL} 
\eeq
and
\beq
S_4(R)\sigma_R^2 \, \simeq \, \frac{216}{25}g_{NL} {\cal{P}}_{\zeta}(1/R)\simeq\frac{4.3}{25}\times 10^{-7}g_{NL}~,
\eeq
where we used the approximations $n_s=1$ and $A_s=2.4\times 10^{-9}$. This analytical analysis 
shows that only primordial non Gaussianities with magnitudes $f_{NL} \geq 10^2$ and 
$g_{NL} \geq 10^6$ yield an important contribution to the modified halo mass function 
of Eq.(\ref{halototal}).
 
Substituting Eq. (\ref{S34}) into the modified halo mass function Eq. (\ref{halototal}) 
we can compute the mass of nonlinear halos with a mean separation $d$ as a function
of redshift. Having such haloes is a necessary condition for the formation of 
super-massive black holes. Since there is evidence that every galaxy harbors a
SMBH, we are interested in the value $d = d_{gal}$ corresponding to the mean
comoving separation of current galaxies 
\bea
\label{condition}
d_{gal}^3 \  M \ \frac{dn}{dM}(M,z) \, = \, 1 \, .
\eea
By using Eq. (\ref{condition}) and the CAMB code \cite{CAMB} we can plot the diagram 
of dark matter mass of halos with mean separation $d_{gal}$ 
as a function of redshift for different values of the
non-Gaussianity parameters $f_{NL}$ and $g_{NL}$ and assuming equilateral shape.
The results are shown in Figure 1 (see also \cite{Reed} for an analysis of the halo
mass function for Gaussian fluctuations).

Our results show that the effect of primordial non-Gaussianities given
by $f_{NL}$ and $g_{NL}$ with values of these parameters consistent with the current 
observational bounds cannot change the number distribution of high redshift
nonlinear dark matter halos in a way which will have an important impact
on the formation of super-massive black holes at high redshifts.
To have an important effect at a redshift $z=20$ values of the non-Gaussianity 
parameters larger than $f_{NL} \geq 10^2$ or $g_{NL} \geq 10^6$ would be
required.

\begin{widetext}

\begin{figure}
\begin{center}
\includegraphics[trim = 0.2in 2.0in 0.02in 2.0in, clip, scale=0.45]{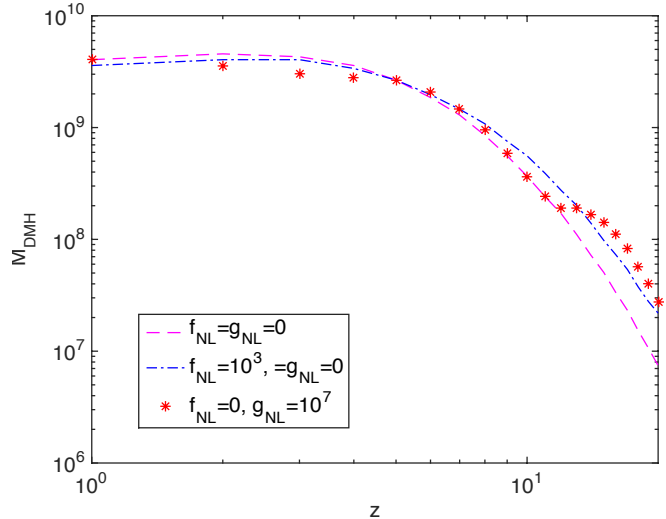}
\caption{Nonlinear dark matter halo mass (vertical axis, in solar mass units) as a function
of redshift (horizontal axis, up to $z=20$) obtained by using the CAMB \cite{CAMB} code for 
different values of the non-Gaussianity parameters $f_{NL}$ and $g_{NL}$ and
assuming equilateral shape. The mass shown is the mass which is turning
nonlinear on a co-moving length scales which corresponds to the separation 
$d_{gal}=1 {\rm Mpc}$ of galaxies. We have made use of the value 
$\frac{t_0}{G}=10^{23} M_{\odot}$. We have shown curves corresponding to the lowest
values of $f_{NL}$ and $g_{NL}$ for which the non-Gaussianities have an
important effect.}
\end{center}
\end{figure}

\end{widetext}

\section{Conclusions}

In the present paper we have studied whether introducing primordial
non-Gaussianities in the density field in terms of skewness and
kurtosis parameters $f_{NL}$ and $g_{NL}$ could have a large impact
on the number density of nonlinear dark matter haloes at high
redshifts. We found that values of $f_{NL}$ and $g_{NL}$ much larger
than the current observational bounds are required in order to change
the predictions appreciably. Hence, density distributions which
are Gaussian modulo skewness and kutosis parameters $f_{NL}$ and $g_{NL}$
cannot have a big impact on the formation of high redshift
super-massive black holes. This result contrasts with the large
effect which cosmic string loops can have \cite{Robert} at high
redshift. In the case of cosmic strings, the distribution is
intrinsically non-Gaussian and not well described at all by
a Gaussian process plus skewness and kurtosis.

\section{Acknowledgments}

We wish to thank Gil Holder and J. Quintin for valuable discussions. 
RB is supported by an NSERC Discovery Grant, and by funds from the 
Canada Research Chair program. 

\bibliography{Databaze}

\begin{thebibliography}{99}

\bibitem{Alexander}
D.~M.~Alexander and R.~C.~Hickox, 
``What Drives the Growth of Black Holes?,"
 New Astron. Rev. {\bf 56}, 93 (2012) 
 [arXiv:1112.1949/astro-ph].
  
\bibitem{Wu}
Xue-Bing.~Wu et al.,
 ``An ultraluminous quasar with a twelve-billion- solar-mass black hole at redshift 6.30", 
 Nature {\bf 518}, 512 (2015).

\bibitem{Fan}
X.~Fan et al. [SDSS Collaboration], 
``A Survey of $z > 5.7$ quasars in the Sloan Digital Sky Survey. 2. Discovery of three additional 
quasars at $z > 6$," 
Astron. J. {\bf 125}, 1649 (2003) 
[arXiv:0301135/astro-ph].
 
\bibitem{Jiang}
L.~Jiang, X.~Fan, M.~Vestergaard, J.~D.~Kurk, F.~Walter, B.~C.~Kelly and M.~A.~Strauss, 
``Gemini Near-infrared Spectroscopy of Luminous $z \sim 6$ Quasars: Chemical
Abundances, Black Hole Masses, and MgII Absorption," 
Astron. J. {\bf 134}, 1150 (2007) 
[arXiv:0707.1663/astro-ph].
 
\bibitem{Willott}
C.~J.~Willott, P.~Delorme, C.~Reyle, L.~Albert, J.~Bergeron, D.~Crampton, X.~Delfosse and 
T.~Forveille {\it et al.},
  ``The Canada-France High-z Quasar Survey: nine new quasars and the luminosity function at redshift 6,''
  Astron.\ J.\  {\bf 139}, 906 (2010)
  [arXiv:0912.0281 [astro-ph.CO]].

\bibitem{Banados}  
  E.~Banados, B.~P.~Venemans, E.~Morganson, R.~Decarli, F.~Walter, K.~C.~Chambers, H.~W.~Rix and E.~P.~Farina {\it et al.},
  ``Discovery of eight $z \sim 6$ quasars from Pan-STARRS1,''
  Astron.\ J.\  {\bf 148}, 14 (2014)
  [arXiv:1405.3986 [astro-ph.GA]].
  
\bibitem{Marconi}
A.~Marconi, G.~Risaliti, R.~Gilli, L.~K.~Hunt, R.~Maiolino and M.~Salvati, 
``Local supermassive black holes, relics of active galactic nuclei and the x-ray background," 
Mon. Not. Roy. Astron. Soc. {\bf 351}, 169 (2004) 
[arXiv:0311619/astro-ph].

\bibitem{Volonteri-Stark}
M.~Volonteri and D.~P.~Stark, 
``Assessing the redshift evolution of massive black holes and their hosts," 
Mon. Not. Roy. Astron. Soc. {\bf 417}, 2085 (2011) 
[arXiv:1107.1946/astro-ph];\\
R.~Salvaterra, F.~Haardt, M.~Volonteri and A.~Moretti, 
``Limits on the high redshift growth of massive black holes," 
Astron. Astrophys. {\bf 545}, L 6 (2012)
[arXiv:1209.1095/astro-ph].

\bibitem{Menou}
K.~Menou, Z.~Haiman and V.~K.~Narayanan, 
``The Merger History of Supermassive Black Holes in Galaxies," 
Astrophys. J. {\bf 558}, 535 (2001) [astro-ph/0101196].

\bibitem{Pollack}
J.~Pollack, D.~N.~Spergel and P.~J.~Steinhardt, 
``Supermassive Black Holes from Ultra-Strongly Self-Interacting Dark Matter,"  
Astrophys.\ J.\  {\bf 804}, no. 2, 131 (2015)
  [arXiv:1501.00017 [astro-ph.CO]].

\bibitem{micic}
M.~Micic, K.~Holley-Bockelmann,  S.~Sigurdsson, L.~Rubbo, 
``Black Hole Growth from Cosmological N-body Simulations," 
Mon. Not. R. Astron. Soc. {\bf 414}, 1127–1144 (2011)  
[arXiv:0805.3154 /astro-ph].

\bibitem{Begelman}
M.~C.~Begelman, M.~Volonteri and M.~J.~Rees,
 ``Formation of Supermassive Black Holes by Direct Collapse in Pregalactic Halos," 
 Mon. Not. Roy. Astron. Soc. {\bf 370}, 289-298, (2006) 
 [arXiv:astro-ph/0602363].

\bibitem{Bromm}
V.~Bromm, 
``The First Stars and Galaxies - Basic Principles," 
[arXiv:1203.3824/astro-ph.CO].

\bibitem{Volonteri}
M.~Volonteri,  
``Formation of Supermassive Black Holes," 
Astron. Astrophys. Rev. {\bf 18}, 279 (2010) 
[arXiv:1003.4404 /astro-ph].
  
\bibitem{Johnson}
J.~L.~Johnson, D.~J.~Whalen, H.~Li, D. E.~Holz, 
``Supermassive Seeds for Supermassive Black Holes," 
[arXiv:1211.0548 /astro-ph.CO].

\bibitem{Robert}
S.~F.~Bramberger, R.~H.~Brandenberger, P.~Jreidini, J.~Quintin, 
``Cosmic String Loops as the Seeds of Super-Massive Black Holes"
 JCAP {\bf 1506}, no. 06, 007 (2015)
  [arXiv:1503.02317 [astro-ph.CO]].

\bibitem{non-minimal}
Y.~Wang,
  ``Inflation, Cosmic Perturbations and Non-Gaussianities,''
  Commun.\ Theor.\ Phys.\  {\bf 62}, 109 (2014)
  [arXiv:1303.1523 [hep-th]].
  
\bibitem{Ekp}
  J.~Khoury, B.~A.~Ovrut, P.~J.~Steinhardt and N.~Turok,
  ``The Ekpyrotic universe: Colliding branes and the origin of the hot big bang,''
  Phys.\ Rev.\ D {\bf 64}, 123522 (2001)
  [hep-th/0103239];\\
J.~L.~Lehners and P.~J.~Steinhardt,
  ``Non-Gaussian density fluctuations from entropically generated curvature perturbations in Ekpyrotic models,''
  Phys.\ Rev.\ D {\bf 77}, 063533 (2008)
  [Phys.\ Rev.\ D {\bf 79}, 129903 (2009)]
  [arXiv:0712.3779 [hep-th]].
  
\bibitem{Wands}
  D.~Wands,
  ``Duality invariance of cosmological perturbation spectra,''
  Phys.\ Rev.\ D {\bf 60}, 023507 (1999)
  [gr-qc/9809062];\\
  F.~Finelli and R.~Brandenberger,
  ``On the generation of a scale invariant spectrum of adiabatic fluctuations in cosmological models with a contracting phase,''
  Phys.\ Rev.\ D {\bf 65}, 103522 (2002)
  [hep-th/0112249];\\
   Y.~F.~Cai, W.~Xue, R.~Brandenberger and X.~Zhang,
  ``Non-Gaussianity in a Matter Bounce,''
  JCAP {\bf 0905}, 011 (2009)
  [arXiv:0903.0631 [astro-ph.CO]].
  
\bibitem{Salopek}
D.~S.~Salopek and J.~R.~Bond, 
``Nonlinear evolution of long wavelength metric fluctuations in inflationary models," 
Phys. Rev. D {\bf 42},  3936–3962 (1990).

\bibitem{Wang}
L.~Verde, L.~M.~Wang, A.~Heavens, and M.~Kamionkowski, 
``Large-scale structure, the cosmic microwave background, and primordial non-gaussianity," 
Mon. Not. Roy. Astron. Soc. {\bf 313}, L 141 (2000), 
[astro-ph/9906301].

\bibitem{Silk}
S.~Yokoyama, N.~Sugiyama, S.~Zaroubi and J.~Silk,
  ``Modification of the halo mass function by kurtosis associated with primordial non-Gaussianity,''
  Mon.\ Not.\ Roy.\ Astron.\ Soc.\  {\bf 417}, 1074 (2011)
  [arXiv:1103.2586 [astro-ph.CO]].

\bibitem{Desjacques}
V.~Desjacques, U.~Seljak, I.~T.~Iliev, 
``Scale-dependent bias induced by local non-Gaussianity: A comparison to N-body simulations,"
 Mon.\ Not.\ Roy.\ Astron.\ Soc.\  {\bf 396}, 85 (2009)
  [arXiv:0811.2748 [astro-ph]].

\bibitem{Smith}
M.~LoVerde, K.~M.~Smith, 
``The Non-Gaussian Halo Mass Function with $f_{NL}$, $g_{NL}$ and $\tau_{NL}$," 
JCAP {\bf 08} 003 (2011)  
[arXiv:1102.1439 / astro-ph].

\bibitem{Maggiore}
M.~Maggiore, A.~Riotto,
``The Halo Mass Function from Excursion Set Theory. III. Non-Gaussian Fluctuations,"  
Astrophys. J. {\bf 717}, 526-541, (2010) [
arXiv:0903.1251/astro-ph].

\bibitem{PlanckCollaboration}
P.~A.~R.~Ade et al. [Planck Collaboration], 
``Planck 2015 results. XVII. Constraints on primordial non-Gaussianity" 
[arXiv:1502.01592/astro-ph.CO]

\bibitem{Amico}
G.~D'Amico, M.~Musso, J.~Noreña, A.~Paranjape,  
``An Improved Calculation of the Non-Gaussian Halo Mass Function", 
JCAP {\bf 02} 001 (2011) 
[arXiv:1005.1203 /astro-ph.CO].

\bibitem{structurebook}
H.~Mo , F.~van den Bosch, S.~White, 
``Galaxy Formation and Evolution," 
(Cambridge University Press, Cambridge, 2010).

\bibitem{PS}
W.~H.~Press and P.~Schechter,  
``Formation of galaxies and clusters of galaxies by selfsimilar gravitational condensation," 
Astrophys. J. {\bf 187}, 425 (1974).

\bibitem{Sheth}
R.~K.~Sheth and G.~Tormen, 
``Large scale bias and the peak background split," 
Mon. Not. Roy. Astron. Soc. {\bf 308}, 119 (1999) 
[astro-ph/9901122].

\bibitem{Matarrese}
S.~Matarrese, L.~Verde, and R.~Jimenez, 
``The abundance of high-redshift objects as a probe of non- gaussian initial conditions", 
Astrophys. J. {\bf 541}, 10 (2000) 
[astro-ph/0001366].

\bibitem{LoVerde}
M.~LoVerde, A.~Miller, S.~Shandera, L.~Verde, 
``Effects of Scale-Dependent Non-Gaussianity on Cosmological Structures", 
JCAP. {\bf 0804} 014 (2008).

\bibitem{CAMB}
A.~Lewis, A.~Challinor and A.~Lasenby, 
``Efficient Computation of CMB anisotropies in closed FRW models", 
Astrophys. J. {\bf 538}, 473 (2000) [astro-ph/9911177].

\bibitem{Reed}
D.~Reed, R.~Bower, C.~Frenk, A.~Jenkins, T.~Theuns, 
``The halo mass function from the dark ages through the present day," 
Mon. Not. Roy. Astron. Soc. {\bf 374}, 2-15, 2007  
[arXiv:0607150 /astro-ph].

\end{thebibliography}

\end{document}